\documentclass[aps,twocolumn,floatfix,showpacs]{revtex4}
\usepackage{graphics}

\newcommand{\ep}{\epsilon}
\newcommand{\bea}{\begin{eqnarray}}
\newcommand{\beq}{\begin{equation}}
\newcommand{\eea}{\end{eqnarray}}
\newcommand{\eeq}{\end{equation}}

\begin{document}
\title{Spectral Singularities and Zero Energy Bound States}
\author{W.\ D.\ Heiss$^{1}$ and R.\ G.\ Nazmitdinov$^{2,3}$}    
\affiliation{$^{1}$National Institute for Theoretical Physics, \\
Stellenbosch Institute for Advanced Study, \\ 
and Institute of Theoretical Physics,
University of Stellenbosch, 7602 Matieland, South Africa\\
$^{2}$Department de F{\'\i}sica,
Universitat de les Illes Balears, E-07122 Palma de Mallorca, Spain\\
$^{3}$ Bogoliubov Laboratory of Theoretical Physics,
Joint Institute for Nuclear Research, 141980 Dubna, Russia}

\begin{abstract}
Single particle scattering around zero energy is re-analysed in view 
of recent experiments with ultra-cold atoms, nano-structures and nuclei far from 
the stability valley. For non-zero orbital angular momentum the low energy scattering 
cross section exhibits dramatic changes depending on the occurrence of either a 
near resonance or a bound state or the situation in between, that is a bound state 
at zero energy. Such state is singular in that it has an infinite scattering length, 
behaves for the eigenvalues but not for the eigenfunctions as an exceptional point 
and has no pole in the scattering function. These results should be observable 
whenever the interaction or scattering length can  be controlled. 
\end{abstract}

\pacs{03.65.Nk,34.50.-s}

\maketitle

\section{Introduction}
\label{intro}
Experimental techniques are nowadays capable to discern subtle phenomena in a 
great variety of fields of physics. 	 
Considerable progress has been made in controlling the interaction of
trapped fermions providing a unique opportunity for the study of different
states of the same atomic system under variation of the atomic interaction.
This is achieved by means of Feshbach resonance techniques \cite{BEC} 
using an external magnetic field; the method allows changing the
scattering length in a wide range between negative and positive values.
In close vicinity of the resonance or a molecular state (a halo dimer) the scattering
length $a$ is very large and changes sign when going from the resonance to the 
molecular state. 

A Bose-Einstein condensate of neutral atoms with induced electromagnetic 
attractive ($1/r$) interaction has been discussed recently as another system 
allowing a tunable interaction \cite{wun2}. The Gross-Pitaevskii equations
describing this system at a large critical negative scattering length yield the 
effective absorbing potential. This critical value where the onset of
the collapse of the condensate occurs could be interpreted as    
a transition point  from separate atoms to the formation 
of molecules or clusters \cite{wun}. In optics, using media with
complex refractive index \cite{guo,makr}, an abrupt phase transition has been
demonstrated. It is associated with the appearance of non-orthogonal super-modes 
near spectral singularities, so-called exceptional points (EP) \cite{ber1,he}.
Nano-structure devices provide another example, where the presence of a quasi-bound state
- resonantly interacting with the continuum of scattering states - leads to a Fano-Feshbach
resonance at specific system parameters \cite{flach}.
All these phenomena can be explained by an effective theory 
of open systems described by a non-Hermitian Hamiltonian,
where complex eigenstates can have square root branch points,
i.e.~EPs.  

We recall that an EP is a singularity of a non-Hermitian Hamiltonian 
where eigenvalues and eigenstates coalesce. The properties of EPs were studied
in earlier experiments using micro-wave cavities \cite{dembo}.
In many cases these singularities produce 
dramatic deviations from the traditional resonance behaviour,
for instance a particularly strong
dependence on the interaction strength (or scattering length)
(see e.g. \cite{HN10,H10} and papers quoted therein).
Even in nuclear physics where an experimental manipulation of, say,
the scattering length appears impossible, an understanding of 
properties of weakly bound nuclei 
in terms of these singularities is being approached \cite{naza}.
Also, using a two-state model it was demonstrated that
a non-Hermitian Hamiltonin can generate the binding of unstable 
states \cite{zel}; the model mimics the properties of {\it halo} nuclei
which form the boundaries for a nuclear valley of stability. 
A kind of unification of nuclear structure and reactions based 
on an effective non-Hermitian Hamiltonian has been suggested 
to understand the transition of the nuclear chart from unbound to
bound limit and {\it vice versa} \cite{naza1}. 

The focus of the present paper lies on the singular behaviour of the energy eigenstates
when the interaction (or scattering length) is varied around a bound state at zero
energy. For $l>0$ the energies behave as if the zero energy bound state was an EP;
yet the eigenfunctions and thus the scattering matrix do not share the singular
behaviour. Nevertheless, the effect upon the cross section is rather dramatic,
especially for $l=1$. There the low energy cross section 'snaps' from an 
$\sim E^2$ behaviour to a $\sim E^0={\it finite\; constant}$ behaviour when a bound state
and an antibound state coalesce at $E=0$ (or likewise when a resonant state
moves to $E=0$).

\section{Resum\'e of known facts}
\label{sec:1}
It is well known \cite{newton} that the single particle scattering problem
with a radial potential that admits bound states can - for angular momentum larger 
than zero - have resonances near to zero energy. By increasing the (attractive) 
potential strength this resonance state evolves via a zero energy eigenstate to 
a bound state. Actually, both the resonance and the bound state appear as a pole 
of the scattering function in the complex $k$-plane ($k\sim \sqrt{E}$ with $E$ 
the energy). In fact, a resonance gives rise to two poles in the lower $k$-plane
that are symmetrically situated with respect to the imaginary $k$-axis.
When increasing the potential strength the two resonance poles in the lower $k$-plane 
move toward $k=0$ where they {\it coalesce}
and then continue moving away at right angle in opposite directions along the imaginary 
$k$-axis. The poles on the positive and negative imaginary $k$-axis correspond to 
a bound and (usually denoted as) anti-bound state, respectively. Note that the 
wave functions associated with the resonances and the anti-bound state
increase exponentially at large distance; they are the Gamow states \cite{max}.
Although these facts are well known, the aspect of the {\it coalescence} at 
$k=E=0$ for specific potential parameters has - to the best of our knowledge - 
not been sufficiently appreciated. We use the term {\it coalescence} as the merging 
of the two eigenvalues does not give rise to the usual degeneracy being characterised 
by two independent eigenfunctions. The type of singularity encountered here appears
to be an EP for the eigenvalues. It therefore produces physically dramatic effects. 
However, as shown below, when looking at the eigenstates and the scattering function 
the behaviour is remarkably different from a genuine EP as it occurs generically 
in finite dimensional matrix problems \cite{gun}.

Without specifying the potential the wave functions are not known
except for their asymptotic behaviour.
For potentials vanishing faster than $r^{-2}$ at large distances, the
asymptotic behaviour of the zero energy bound state wave function is
for orbital angular momentum $l$
\beq
\psi_l(r)\sim \frac{1}{r^{l+1}}.
\eeq
General statements are known for the behaviour of the scattering length, the phase
shift and the cross section at low energies. For non-zero energy eigenvalues
the scattering length $a_l$ is defined by the expansion
\beq
\exp 2i\delta_l=1+i a_lk^{2l+1}+O(k^{2k+2})
\eeq
where $\delta _l$ is the scattering phase shift and $|\exp 2i\delta_l-1|^2/k^2$ 
is proportional to the cross section. It is further known that $a_l>0$ for the 
case of a low energy resonance 
and that $a_l$ tends to infinity for a zero energy eigenvalue. Similarly, it is
known that $a_l<0$ when a bound state has just emerged. In other
words, the scattering length has a first order pole when an eigenvalue
occurs at $E=0$.

What happens to Eq.(2) in this latter situation? We now turn to this question.

\begin{figure}
\resizebox{0.30\textwidth}{!}{%
\includegraphics{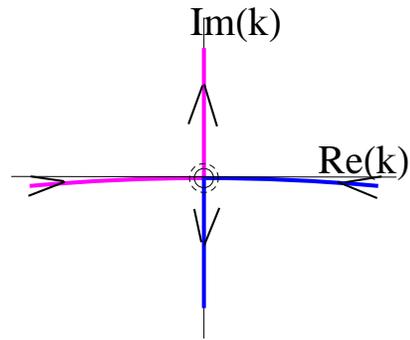}
}
\caption{Trajectories in the $k$-plane of the two resonances (horizontal curves) 
when $\ep <0$ tends to zero. At $\ep =0$ the two eigenvalues coalesce at $k=0$ and 
move for increasing $\ep >0$ in opposite directions along the imaginary $k$-axis.}
\label{traj}
\end{figure}

\section{Analytic treatment}
\label{sec:2}
The motion of the poles of the scattering function, that is of the eigenvalues of the 
Schr\"odinger equation using outgoing wave boundary conditions, seem to indicate
the typical signature of an EP at $E=k=0$: a square root singularity in the potential 
strength. Given the potential strength $v_0$ that produces a bound state at $k=0$ 
for angular momentum $l>0$, changing the potential to
$v_0+\ep$ yields the new eigenvalues as
\beq
k_{1,2}=\sum _{n=1}^{\infty} c_n^{(1,2)}\sqrt{\ep}\,^n 
\eeq
with finite radius of convergence. The labels 1 and 2 refer to the two resonances 
for $\ep <0$; accordingly, for $\ep >0,\, E_1=k_1^2$ and $E_2=k_2^2$ are the bound 
and anti-bound states energies, respectively. Note that $E_{1,2}<0$ for $\ep >0$
while $E_{1,2}$ are complex for $\ep <0$.
The occurrence of the square root in (3) is clearly reminiscent 
of an EP and also clearly signals the sprouting out of the two energies in 
different directions depending on the sign of $\ep $.
We stress that the square root behaviour in (3) refers to the
{\it eigenvalues as a function of the potential strength}. Associating
for the scattering function
 - as usually - the upper/lower $k$-plane with the first/second sheet of the energy
plane, respectively, we clearly see that an emerging bound state - a
pole of the scattering function in the first energy sheet at $E_1<0$ -
will always have an antibound state - a pole in the second sheet at $E_2<0$ - as a partner.

We illustrate this result explicitly for a square well of width
$\pi$ and $l=1$. In this case the $p$-wave bound state at $E=k=0$
appears for $v_0=1$. One finds the explicit expansions
\bea
k_1&=&+i(\frac{\sqrt{\ep}}{\sqrt{3}}+\frac {\pi}{9}\ep+O(\ep ^{3/2})) \\
k_2&=&-i(\frac{\sqrt{\ep}}{\sqrt{3}}-\frac {\pi}{9}\ep+O(\ep ^{3/2})) 
\eea
indicating nicely the motion of the levels (the poles of $\exp
2i\delta_1$) when $\ep$ changes from small negative to positive values
as illustrated in Fig.1.

In a similar way one finds for $\ep =0$ the expansion of the
scattering function around $k =0$
\beq
\exp 2i\delta_1=1-i\frac{4}{3}\pi k+O(k^2) 
\eeq
while (2) remains valid for $\ep \ne 0$. We note that for $l>1$ the leading
power of $\ep $ in (4) and (5) persists while that of (6) is to be replaced by $k^{2l-1}$.

\section{Singularity and observable consequences}
\label{sec:3}
This last result invokes dramatic consequence for low energy scattering.
It is illustrated in Fig.2, where the cross section
\begin{figure}
\resizebox{0.35\textwidth}{!}{%
\includegraphics{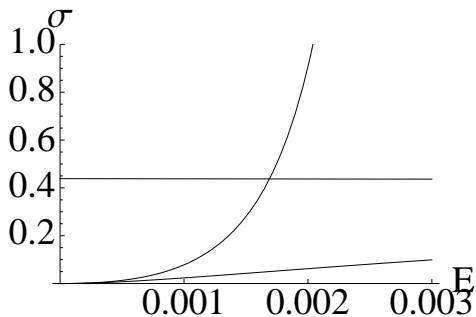}
}
\caption{Cross sections versus energy (in arbitrary units) for $\ep =0$ 
(straight line), $\ep <0$ (sharp rising curve) and $\ep >0$ (lower curve).}
\end{figure}
- being proprtional to $|1-\exp (2i\delta_1|^2/k^2$ - is drawn for $\ep =0$ 
(straight line), for $\ep =-10^{-2}$ (sharp rising curve) and $\ep=+10^{-2}$ 
(lower curve). Only when $\ep =0$ is the low energy cross section a constant in the
energy (we simply consider the cross section integrated over the angles). As soon as
$\ep \ne 0$ the cross section starts at zero and rises with the quadratic power 
irrespective of the sign of $\ep $. This dramatic change at low energy scattering 
should be observable whenever the potential or scattering length can be controlled 
as in the experiments discussed in the introduction. The cross sections for the
different signs of $\ep \ne 0$ are almost identical for small values of energy. 
The onset of the sharp rise for $\ep <0$ with increasing energy is due to the 
resonance generated by the slightly less attractive potential. This rise depends 
directly upon the magnitude of the negative value of $\ep $: the smaller $|\ep |$ 
the smaller the resonance energy, that is the nearer to zero the occurrence of 
the sharp rise; yet the initial rise still remains $\sim E^2$.
\begin{figure}[b]
\resizebox{0.35\textwidth}{!}{%
\includegraphics{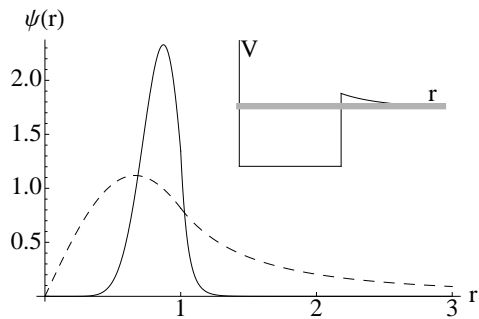}
}
\caption{Normalised zero energy bound state wave functions for $l=1$ (dashed) 
and $l=9$ (solid) for potentials of different depths; 
here the width of the potential has been chosen unity implying about a factor 
three for the quotient of the two potential depths. The inset illustrates 
schematically the effective potential.}
\end{figure}

This sudden switch in the low energy behaviour is a manifestation of the specific
{\it singularity} for the zero energy bound state. Yet, while the spectrum appears 
to have the signature of an EP, neither the wave function and hence the scattering 
function displays the typical characteristics of an EP. In fact, while the 
resonances and the bound and anti-bound states are represented by a first order 
pole of $\exp 2i\delta_l$, the two residues (the spectroscopic factors) conspire 
in the limit $\ep \to 0$ such that the pole {\it does not occur }
for the bound state at zero energy. In other words, the spectroscopic
factors vanish when the zero energy bound state is approached.
This is in sharp contrast to the behaviour of a genuine EP in finite
matrix models or for the coalescence of two resonances at a complex energy. 
There the scattering function (or Green's function) has a pole of 
second order \cite{mondra,HN10}. The double pole is related to the vanishing norm
of the coalescing eigenfunctions at the EP (sometimes referred to as 
self-orthogonality \cite{mois}). In other words, there the spectroscopic factors 
become large in close vicinity of an EP \cite{naza1}.
In the present case, the zero energy bound state wave function is normalisable. 
Therefore, even though the energies appear to behave like the levels at an EP 
including the singular behaviour, the eigenfunctions do not. Note that an EP 
cannot appear for a self-adjoint Hamiltonian, and the zero energy bound state problem
falls into this class.

We emphasise that our findings are expected to be valid irrespective of the 
particular form of the radial potential as long as it falls off faster than 
$r^{-2}$ and is less singular than $r^{-2}$ at $r=0$. In particular, the results 
are valid for an effective single particle problem describing nucleon-nucleus 
scattering. It is here where our results are expected to have a bearing even in 
nuclear physics.  Note that the bound state at zero energy is ``loosely'' bound as
seen by the mild decrease of the wave function. As stated above the state has 
all characteristics of a Feshbach resonance; in fact the slightest change of 
external parameters (scattering length) may turn it into a proper resonance or 
a weakly bound state. Of course, in nuclear physics the only change of parameters 
is moving along the isotopic or isotonic line. The situation discussed resembles
that of nuclei on the drip line. In connection with halo nuclei we stress that our 
findings are valid for any angular momentum larger than zero. We recall in this 
context that the wave function for a zero energy bound state is - for higher 
partial waves - sharply concentrated at the surface of the binding potential, 
the more so the larger the angular momentum. In Fig.3 we illustrate the zero energy 
bound states for $l=1$ and $l=9$. The inset shows schematically the potential 
indicating the role of the centrifugal part of the potential. The thick grey line 
indicates the energy region in which we concentrate our discussion of such
Feshbach resonance. There the dramatic features for low energy scattering under 
variation of, for instance, the scattering length or the potential depth do occur, 
as has been discussed above in more detail.

In passing we note a mathematical subtlety. Even though the wave function of the 
anti-bound state cannot be normalised (it grows exponentially like the 
resonance states), we may calculate the scalar product of the bound state
and the anti-bound state wave function. The exponential decay of the bound state 
wave function outweighs the exponential growth of the anti-bound state. As a 
result the integral converges and is exactly zero: the bound state is 
``orthogonal'' upon the anti-bound state. It looks like a contradiction as the 
anti-bound- and bound state become the more alike the smaller $\ep >0$. The apparent
contradiction is due to the non-uniform behaviour related to the
singular behaviour: the exponential growth of the anti-bound state
is shifted to infinity when $\ep \to 0$; if the limit is taken before integration 
the scalar product does not vanish; it is the integral over the far outside tail 
that brings about the cancellation with the first finite part of the integral. 
Explicit expressions are given in the Appendix.

We suggest that, similar to the Gamow resonance states, the anti-bound wave 
function can have a physical meaning. Matrix elements describing transitions 
containing resonance states are widely used. In a similar vein the anti-bound 
state may be viewed as a resonance at zero frequency with a finite width. With 
experimental techniques available nowadays this could be within reach. 
We recall that the significance of an s-wave anti-bound state is known since 
long for the neutron-neutron system \cite{nn}.

\section{Conclusion}
To summarise, we have shown that at low energy single particle scattering for 
angular momentum larger than zero there is a 
dramatic difference for the cross section between the special situation of a zero
energy bound state and the existence of either a resonance or a bound state at 
finite energy. For $l=1$ the cross section is constant in the case of a zero 
energy bound state, while it obeys a quadratic energy dependence for low energy 
eigenstates with nonzero energy (resonance or bound state). While these results 
have been known in principle \cite{rau}, the connection to a specific singularity 
has - to the best of our knowledge - not been made. In fact, it is remarkable that 
the scattering function has no pole at a {\it resonance at zero frequency} whereas
the cross section remains finite. We note
that for $l>1$ the pattern for the cross section translates into a $\sim E^{2l-2}$ 
behaviour for the zero energy bound state and into a $\sim E^{2l}$ behaviour for 
the non-zero eigenvalues. We think that our findings 
are relevant, for instance, for experiments with trapped ultra-cold atoms in 
either fermion and boson gases, where the scattering length can be controlled 
near the threshold energy. We believe that the insights gained also have a bearing 
for the understanding of  nuclei around the drip line where weakly bound states 
and larger angular momenta are relevant.

\section*{Acknowledgements}
This work is partly supported by JINR-SA Agreement on scientific collaboration, 
by Grant No. FIS2008-00781/FIS (Spain), and by
the RFBR Grant No.11-02-00086 (Russia). 
\section*{Appendix}
For $l=1$ the asymptotic behaviour of the bound state wave function is
for large distances $r$
$$
|\psi _{\rm bound}\rangle \sim \frac{e^{-k_b r} (1+k_b r)}{r^2}
$$
and for the antibound state 
$$
|\psi _{\rm a-bound}\rangle \sim \frac{e^{+k_a r} (1-k_a r)}{r^2}
$$
with $k_b>k_a>0$. Obviously the two functions become identical for
$k_b\to 0$. We consider the integral giving the scalar product.
The tail end section (being the range from the zero of 
$|\psi _{\rm a-bound}\rangle $ at $r=1/k_a$
to infinity) reads
$$
\int _{1/k_a}^\infty \frac{e^{-k_b r} (1+k_b r)}{r^2} \frac{e^{+k_a r}
  (1-k_a r)}{r^2} r^2 {\rm d}r = 
\frac {\exp (1-\frac{k_b}{k_a}) k_a^2} {k_b-k_a}.
$$
It remains finite in the limit
$k_b\to 0$ since $k_a^2/(k_b-k_a)$ remains finite  (implying $k_a\to 0$) as seen from
(3); an example is given in (4) and (5) when the limits $\ep \to 0$ are taken
for $k_b=-i k_1$ and $k_a=i k_2$. Note that the integral
above would vanish if the limit is taken before integration. For
$k_a\ne 0$ it is the finite value of the tail end integral that
cancels against the first part of the scalar product integration thus
causing ``orthogonality'' of $|\psi _{\rm bound}\rangle $ and
$|\psi _{\rm a-bound}\rangle $ for $k_a\ne 0$. Yet, when $k_b=k_a=0$
the two functions are identical and have a finite norm.

\end{document}